\documentclass[aps,prl,reprint,superscriptaddress]{revtex4-1}

\usepackage{graphicx}
\usepackage{color}
\usepackage{amsmath,amssymb}
\usepackage{bm}
\usepackage{upgreek}
\usepackage{xspace}

\usepackage[colorlinks,urlcolor=blue,citecolor=blue,linkcolor=blue]{hyperref}
\graphicspath{{figures/}}

\newcommand{\uvect}[1]{\hat{\boldsymbol{#1}}\xspace}

\begin{document}

\title{$T_2$-limited sensing of static magnetic fields via fast rotation of quantum spins}

\author{A.~A.~Wood, A.~G.~Aeppli, E. Lilette, Y.~Y.~Fein, A.~Stacey, L. C. L. Hollenberg, R. E. Scholten and A. M. Martin}
\affiliation{School of Physics, University of Melbourne, Victoria 3010, Australia.}

\date{\today}

\begin{abstract}
{Diamond-based quantum magnetometers are more sensitive to oscillating (AC) magnetic fields than static (DC) fields because the crystal impurity-induced ensemble dephasing time $T_2^*$, the relevant sensing time for a DC field, is much shorter than the spin coherence time $T_2$, which determines the sensitivity to AC fields. Here we demonstrate measurement of DC magnetic fields using a physically rotating ensemble of nitrogen-vacancy centres at a precision ultimately limited by $T_2$ rather than $T_2^*$. The rotation period of the diamond is comparable to $T_2$ and the angle between the NV axis and the target magnetic field changes as a function of time, thus upconverting the static magnetic field to an oscillating field in the physically rotating frame. Using spin-echo interferometry of the rotating NV centres, we are able to perform measurements for over a hundred times longer compared to a conventional Ramsey experiment. With modifications our scheme could realise DC sensitivities equivalent to demonstrated NV center AC magnetic field sensitivities of order $0.1$\,nT\,Hz$^{-1/2}$.}

\end{abstract}

\maketitle
\section{Introduction}
Precision sensing of low-frequency and static fields is of interest to applications such as low-field NMR~\cite{budker_optical_2013}, magnetoencephalography~\cite{hamalainen_magnetoencephalography_1993} and the study of magnetic structures~\cite{tetienne_nanoscale_2014, simpson_magneto-optical_2016, tetienne_quantum_2017}, biological samples~\cite{mcguinness_quantum_2011,le_sage_optical_2013} and magnetic anomaly detection \cite{fleig_maritime_2018}.
Most magnetic sensors exhibit reduced sensitivity at low frequency and DC; for example, SQUIDs~\cite{braginski_squid_2004} are subject to noise with a $1/f$ frequency dependence and thus exhibit better sensitivity to AC magnetic fields than low-frequency or DC fields. For quantum sensors based on nitrogen-vacancy (NV) electron spins in diamond~\cite{doherty_nitrogen-vacancy_2013, rondin_magnetometry_2014}, the DC magnetic field sensitivity is limited by the ensemble dephasing time $T_2^*$~\cite{dreau_avoiding_2011} which is determined by phase-incoherent sampling of magnetic or electric fields originating from other spins within the crystal, as well as other factors such as variations of temperature~\cite{acosta_temperature_2010,fang_high-sensitivity_2013}. Inhomogeneous broadening across an ensemble of NV sensors due to crystal strain or magnetic gradients has similar effects~\cite{bauch_ultralong_2018}. Detecting an AC field with frequency $f\sim 1/\tau$ using a spin-echo sequence~\cite{hahn_spin_1950} with interpulse spacing $\tau$ refocuses perturbations caused by low-frequency noise on a shot-by-shot basis, and extends the maximum sensing time to $\tau_\text{max}\sim T_2$, the electron spin coherence time~\cite{taylor_high-sensitivity_2008}. $T_2$ is typically several orders of magnitude larger than $T_2^\ast$, resulting in sensitivity improving by $\sqrt{T_2^\ast/T_2}$, but at the expense of insensitivity to DC magnetic fields. In this work, we demonstrate a quantum magnetometry method based on an ensemble of rotating spin qubits which can detect DC magnetic fields with a sensitivity ultimately limited by $T_2$, rather than $T_2^*$. 

Our technique upconverts the DC magnetic field to AC by rotating the host diamond crystal with a period comparable to $T_2$~\cite{wood_magnetic_2017, wood_quantum_2018} in a way that modulates the coupling between the NV center and the magnetic field to be detected. With the NV crystal axis at an angle $\theta_\text{NV}$ to the rotation axis and a small target DC magnetic field transverse to the rotation axis, the Zeeman shift of the NV is modulated at the rotation frequency in proportion to the DC field. Since the NV Zeeman shift is now time dependent, we can use rotation-synchronized spin-echo AC magnetometry, which refocuses in-crystal noise and extends the maximum sensing time to $T_2 \gg T_2^\ast$. Upconversion using sensor rotation means the field to be detected remains fundamentally DC, and yet allows AC magnetometry with the measurement frequency and phase well-defined and synchronised to the rotation of the diamond. We demonstrate more than a hundredfold increase in DC field sensing time $\tau$, from $\tau = T_2^\ast$ to $\tau \approx T_2$, and a correspondingly enhanced response to DC fields relative to the best conventional Ramsey magnetometry possible with our diamond sample. Our method could bring to DC field sensing with the NV center sensitivities comparable to diamond-based AC magnetometers~\cite{wolf_subpicotesla_2015}.

\section{Magnetic sensing at DC with solid-state spins}

Coherent quantum sensing with solid-state spin systems takes place amongst interactions between the central sensor spin and a surrounding bath of other spins. In bulk diamond, these bath spins are typically either substitutional nitrogen (P1) electron spins~\cite{vlasov_relative_2000, kennedy_long_2003, hanson_coherent_2008} or $^{13}$C nuclear spins~\cite{childress_coherent_2006}. Due to the projective nature of quantum measurement and very low photon collection efficiencies, it is generally necessary to average measurements over integration times well in excess of typical bath correlation timescales. As a result, the perturbations from bath spins are incoherently averaged, leading to dephasing on a timescale of $T_2^\ast$. Here, $T_2^\ast$ refers to dephasing over repeated projective measurements, rather than the conventional NMR definition which typically pertains to inhomogeneous precession in an ensemble of spins~\cite{vandersypen_NMR_2005}. 

A simple solution that improves DC and AC magnetic sensitivity in NV diamond magnetometers consists of using istopically-pure $^{12}$C diamond samples~\cite{balasubramanian_ultralong_2009}. However, for practical magnetometry the high densities of nitrogen required for ensemble densities of NV centres result in $T_2^\ast$ being far lower than that set by the depleted $^{13}$C limit. Driving the bath spins using resonant radiofrequency has been shown to average the perturbative effect and increase $T_2^\ast$ \cite{lange_controlling_2012} though the demonstrated increase remains well below the $T_2$ limit~\cite{bauch_ultralong_2018}. A number of alternative measurement schemes exist to improve NV magnetic sensitivity to low-frequency and DC magnetic fields, which have focused on level-crossing dynamics~\cite{wickenbrock_microwave-free_2016}, ancillary nuclear spins~\cite{ajoy_DC_2016}, improving collection efficiency using IR-absorption magnetometry~\cite{acosta_broadband_2010} and creating an NV-based lasing cavity~\cite{jeske_laser_2016}.

Up-conversion using rotation is a simple alternative that brings the benefits of AC measurement to DC sensing, with the particularly attractive feature that the only required modification is sensor rotation. A similar procedure is demonstrated in Ref.~\cite{schmidt_getmag_2004}, where rotating a SQUID gradiometer was shown to upconvert DC magnetic signals of geophysical origin to the rotation frequency (tens of Hz), away from $1/f$ noise. In Ref.~\cite{hong_coherent_2012}, a magnetic tip attached to a tuning fork resonator was used to rapidly modulate the position of the tip relative to a single NV. The strong field gradients from the tip resulted in modulation of the NV Zeeman shift with the tip movement, and spin-echo magnetometry measured the up-converted DC magnetic field. Up-conversion with sensor rotation as demonstrated in our work relies on the vector properties of the magnetic field, rather than modulating the source of the magnetic field to be measured, and is therefore equally applicable to microscale sensing with rotating single qubits \cite{wood_quantum_2018} and macroscopic sensing with large ensembles of NV sensors. 

\section{Experiment}\label{sec:exp}
\begin{figure*}[t!]
	\centering
		\includegraphics[width=\textwidth]{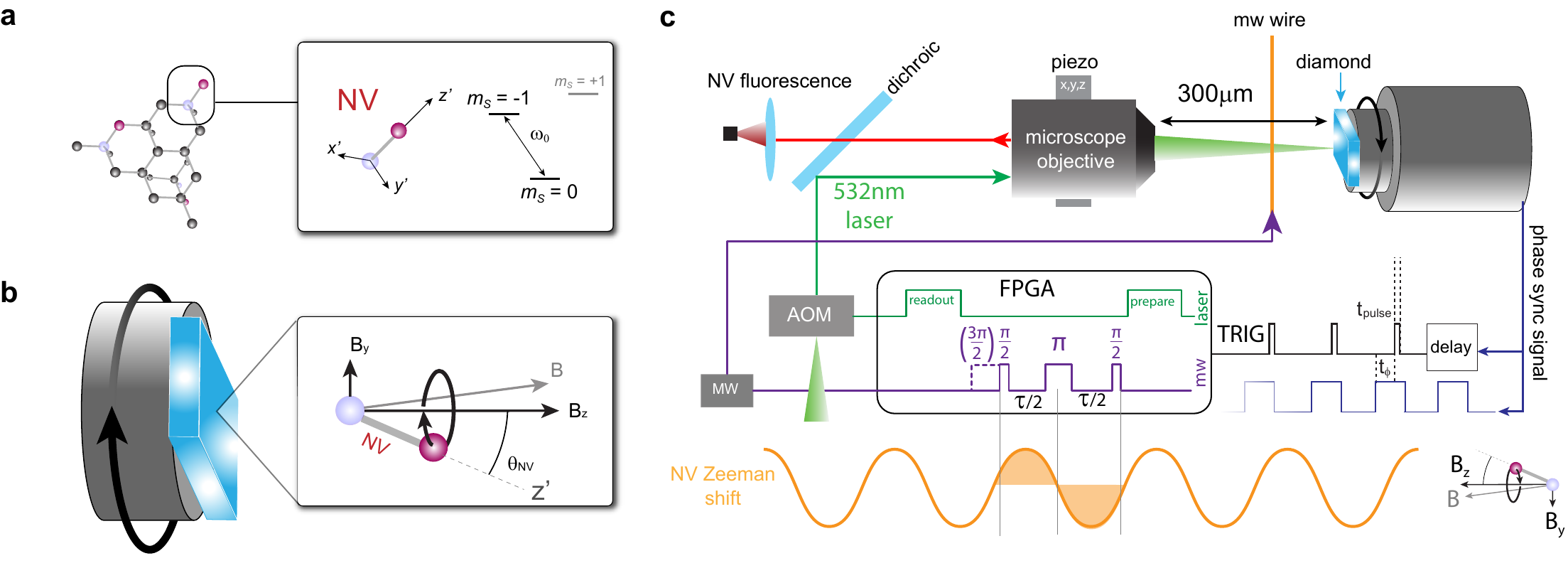}
	\caption{Experimental configuration and rotational up-conversion procedure. (a) Outline of the NV centre in diamond showing coordinate frame and effective two-level system. (b) A diamond containing an ensemble of NV centres is mounted to a high-speed electric motor that rotates the diamond around an axis $z$ at up to $200,000$\,rpm (3.33\,kHz). One of the four NV orientation classes ($z'$) makes an angle of $\theta_\text{NV}\approx4^\circ$ to the rotation axis. A $5.7\,$mT magnetic bias field is aligned parallel to $z$. When a transverse magnetic field $B_y$ is applied, the Zeeman shift of the NV becomes time dependent since it is proportional to ${\bf B}(t)\cdot\uvect{z'}$ in the NV frame. The time-dependent magnetic field can then be measured with a spin-echo pulse sequence performed in the physically rotating frame, with the relevant sensing timescale now determined by $T_2$ rather than the ensemble dephasing time $T_2^\ast$. (c) Experimental setup. The motor controller outputs a pulse synchronous with the rotation which triggers an FPGA pulse generator. The FPGA outputs a spin echo pulse sequence, consisting of laser preparation and readout pulses (separated by one rotation period) and a $\pi/2$-$\pi$-$\pi/2\,\,(3\pi/2)$ microwave pulse sequence with pulse spacing $\tau/2$. Laser light is focused onto the centre of the diamond rotation with a high-NA objective, which also collects the emitted photoluminesence and directs it onto an avalanche photodiode.}
	\label{fig:scematic}
\end{figure*}

The experimental setup and methods are depicted in Fig.~\ref{fig:scematic} and are similar to those described previously \cite{wood_magnetic_2017, wood_quantum_2018}. A diamond containing an ensemble of NV centers is mounted to the spindle of an electric motor that rotates at 200,000\,rpm (3.33\,kHz). In the stationary coordinate system with $z'$ the NV axis, the Hamiltonian is time-independent and given by 
\begin{equation}
H = D_\text{zfs} S_{z'}^2 +\gamma {\bf S}\cdot{\bf B},
\label{eq:ham}
\end{equation}
with the electron gyromagnetic ratio $\gamma = 28\,\text{GHz\,T}^{-1}$, ${\bf S} = (S_{x'}, S_{y'}, S_{z'})$ the spin vector,  $D_\text{zfs} = 2.87\,$GHz the zero field splitting and ${\bf B}$ the magnetic field. When $\gamma|B|\ll D_\text{zfs}$, the Zeeman shift term depends only on the strength and orientation of the magnetic field. If the NV axis is then rotated around $z$ at an angular frequency $\Omega$ in the presence of a magnetic field with, for example, a transverse $y$-component ${\bf B} = (0,B_y, B_z)$, the Zeeman shift becomes time-dependent, and the $m_S = 0\rightarrow m_S = -1$ transition frequency is given by
\begin{align}
\omega(t) = &\, D_\text{zfs} - \gamma\left[B_z\cos\left(\theta_\text{NV}\right)-B_y\sin\left(\theta_\text{NV}\right)\cos\left(\Omega t-\phi_0\right)\right]\nonumber\\
          \equiv &\, \omega_0 +\gamma B_\text{UC}(t),
\label{eq:simp}
\end{align}       
with $\theta_\text{NV}$ the angle between the NV axis and rotation axis, $\phi_0$ some arbitrary initial phase, $\omega_0 = D_\text{zfs}-B_z\cos\left(\theta_\text{NV}\right)$ and $B_\text{UC}(t) = B_y\sin\left(\theta_\text{NV}\right)\cos\left(\Omega t-\phi_0\right)$ the up-converted DC $B_y$ field, oscillating at the rotation frequency. Rotation of the NV axis modulates the Zeeman shift at the rotation rate, and the projection of the DC field parallel to the NV axis $B_y\,\sin\left(\theta_\text{NV}\right)$ yields sensitivity to the transverse field component $B_y$ in the NV frame.

The diamond sample we use is a $(111)$-cut electronic grade sample grown using chemical vapour deposition, containing natural ($1.1\,\%$) abundance $^{13}$C and an NV concentration of $10^{15}\,\text{cm}^{-3}$. Of the four orientation classes present in the diamond, one makes an angle of $3.8^\circ$ to the rotation axis while the other three make angles of approximately $106^\circ,112^\circ$ and $111^\circ$. While the sensitivity to a magnetic field transverse to the rotation axis is proportional to $\sin\,\theta_\text{NV}$, we use the orientation class with $\theta_\text{NV} = 3.8^\circ$. Although three orientation classes make large obtuse angles ($\sin\,\theta\approx 0.93$) to the rotation axis, the requirement to form an addressable two-level system for quantum measurement necessitates a static bias field to break the degeneracy of the $m_S = \pm1$ states of the NV ground state. In the presence of this bias field, which must be parallel to the rotation axis, these three orientation classes are essentially degenerate, and measurement of a particular class using state-dependent photoluminesence becomes essentially impossible. We therefore choose to address the NV orientation class almost parallel to the rotation axis, since this greatly simplifies our measurement protocol. This limitation can be circumvented with diamonds containing single NV centres,  preferentially oriented NVs~\cite{pham_enhanced_2012} or crystals polished to an appropriate angle, as detailed in the Discussion.

In the presence of a $5.7\,$mT magnetic bias field parallel to the rotation axis, the two-level splitting of the $m_S = 0$ and $m_S = -1$ Zeeman states is $2.711\,$GHz and we use microwave pulses resonant with this transition to control the populations and coherences for quantum sensing. A microscope objective mounted on a scanning piezoelectric stage focuses 532\,nm light to a 600\,nm spot and directs red fluorescence emitted by the NV centers onto an avalanche photodiode, in a confocal microscope configuration. A 20\,$\upmu$m diameter copper wire located 100$\,\upmu$m above the diamond surface is used to apply microwave fields. Magnetic fields are applied using a single multi-turn coil coaxial with the motor spindle, behind the diamond. Coil pairs along the $x$ and $y$ axes are used to create the test fields for rotational up-conversion. A 1.0\,mm thick mu-metal shield on the front face of the motor screens the diamond from magnetic fields originating from the pole pieces of the motor.

Optical preparation, readout and microwave state manipulation sequences are synchronized to the rotation of the diamond using a pulse generator triggered by the electric motor phase synchronization signal. The trigger ensures phase synchronicity with the up-converted field, and can be delayed in order to synchronize the interferometric sequence to any particular phase. We operate with the focus of the preparation and readout laser beam positioned as close as possible to the rotation centre of the diamond to reduce the effects of NV motion during quantum state preparation and readout. A 3$\upmu$s laser pulse is applied to prepare the NV ensemble into the $m_S = 0$ state, followed by a $\pi/2 -\pi-\pi/2$ spin-echo microwave pulse sequence with interpulse spacing $\tau/2$. The fluorescence contrast $\mathcal{S}_i$ from $10^5$ or more repetitions of the experiment with a final $\pi/2$ projection pulse is compared to a sequence with the final $\pi/2$ pulse replaced with a $3\pi/2$ pulse in order to compute a normalised spin echo signal $\mathcal{S} = (\mathcal{S}_{\pi/2}-\mathcal{S}_{3\pi/2})/(\mathcal{S}_{\pi/2}+\mathcal{S}_{3\pi/2})$.  

\section{Coherence of rotating qubits} 

\begin{figure}
	\centering
		\includegraphics[width = \columnwidth]{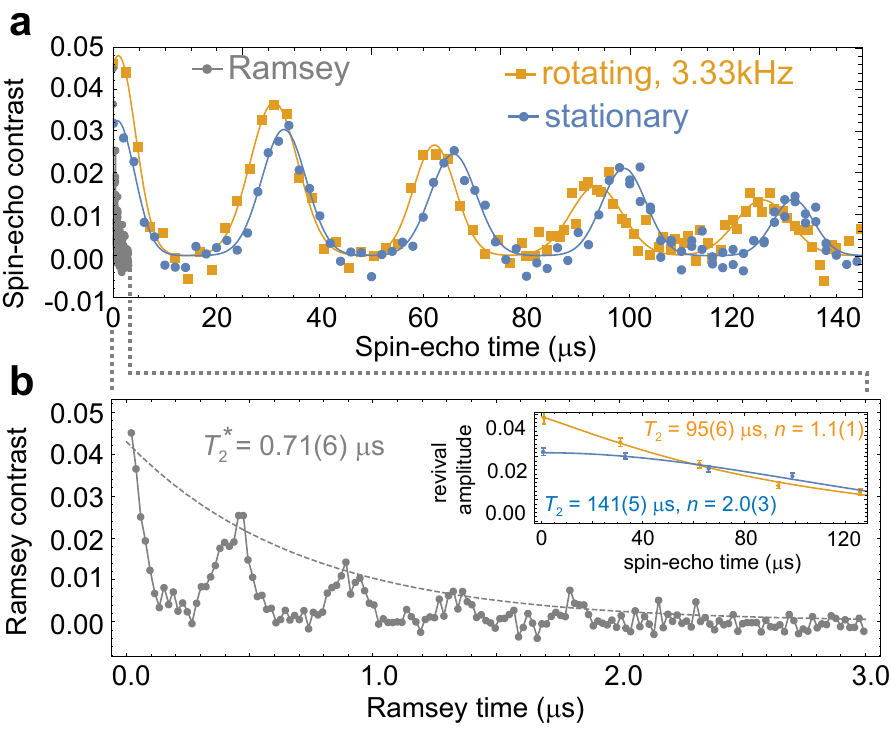}
	\caption{Coherence of rotating NV ensemble. (a) Stationary NV Ramsey data (grey circles) and NV spin-echo signal at $B_z = 5.7\,$mT when stationary (blue circles) and when rotating at 3.33\,kHz (orange squares). A rotationally-induced magnetic pseudo-field adds to $B_z$ for the rotating data, shifting the position of the $^{13}$C revivals. Refer to Fig. \ref{fig:VSWIM}(a) for representative error bounds. (b) Detail of Ramsey contrast showing $T_2^\ast$ decay envelope due to quasi-static ensemble dephasing. The oscillations present in the Ramsey data are due to the $^{14}$N nuclear hyperfine interaction. Inset: for comparison, the decay envelopes of rotating and stationary spin-echo signal, with effective $T_2$ time and decay exponent (see main text) allow for $>100$ times longer interferometric interrogation. Error bars derived from uncertainty in Gaussian fits to spin-echo revivals.}
	\label{fig:revivalsdata}
\end{figure}

Before proceeding to a demonstration of rotational up-conversion magnetic sensing, we briefly examine deleterious effects of physical rotation or up-converted noise by performing a simple experiment: comparing the stationary and rotating spin-echo signals. For diamonds containing $<1$\,ppm N and natural abundance $^{13}$C, the dephasing time is at most $\sim 3\,\upmu$s due to phase-incoherent sampling of the nuclear magnetic dipole fields by the NV during a Ramsey measurement~\cite{rondin_magnetometry_2014}. Weaker dipolar-mediated $^{13}$C spin flip-flops limit the maximum spin-echo coherence time $T_2$ to a few hundred microseconds. Since the spin-echo signal is modulated by the $^{13}$C interaction, measurement contrast is limited to revivals spaced at multiples of twice the nuclear spin Larmor precession period~\cite{childress_coherent_2006}. Ensuring spin-echo contrast exists amounts to satisfying $\tau = j (2\gamma_{13C} B_z)^{-1} <T_2$, with $\gamma_{13C} = 10.7\,$MHz\,T$^{-1}$ and integer $j$. Figure \ref{fig:revivalsdata} shows the stationary Ramsey and spin-echo signals compared to the spin-echo signal with the diamond rotating at 3.33\,kHz, at a common bias magnetic field strength of $B_z =5.7\,$mT. The Ramsey signal decays rapidly, with a characteristic time of $T_2^* = 0.71(6)\,\upmu$s. The spin-echo signals exhibit the characteristic $^{13}$C modulation with an overall decoherence envelope $\exp\left(-(\tau/T_2)^n\right)$ due to nuclear spin bath dynamics~\cite{zhao_decoherence_2012}. The decoherence time $T_2$ and decay exponent $n$ change depending on the strength of the magnetic field~\cite{hall_analytic_2014}. 

An obvious difference between the stationary and rotating spin-echo signals is a shift in the revival times, due to rotation of the diamond. In this experiment, rotation opposes the precession direction of the $^{13}$C nuclear spins, resulting in a $0.31\,$mT magnetic pseudo-field adding to the bias magnetic field for the $^{13}$C spins~\cite{wood_magnetic_2017}. Rotating-NV sensor spin-echo coherence exists well beyond the $T_2^*$ limit with no other substantial difference between rotating and stationary NV spin-echo signals. This demonstrates that no significant noise sources are introduced to the spin-echo measurement due to the physical rotation of the diamond. For the data shown in Fig. \ref{fig:revivalsdata} the spin-echo pulse sequence was not synchronized to the rotation of the diamond and we used a measurement scheme described previously in Ref. \cite{wood_magnetic_2017}. Any misalignment of the bias magnetic field from the rotation axis then manifests as an up-converted field with random phase, which reduces the measured $T_2$. Nevertheless, we observe interferometric contrast of a comparable magnitude to the stationary spin-echo signal in the vicinity of the fourth $^{13}$C revival, near $\tau = 124\,\upmu$s, and use this as the sensing time to demonstrate rotational up-conversion.

\section{DC magnetometry with rotational up-conversion}\label{sec:dcup}
\begin{figure}
	\centering
		\includegraphics{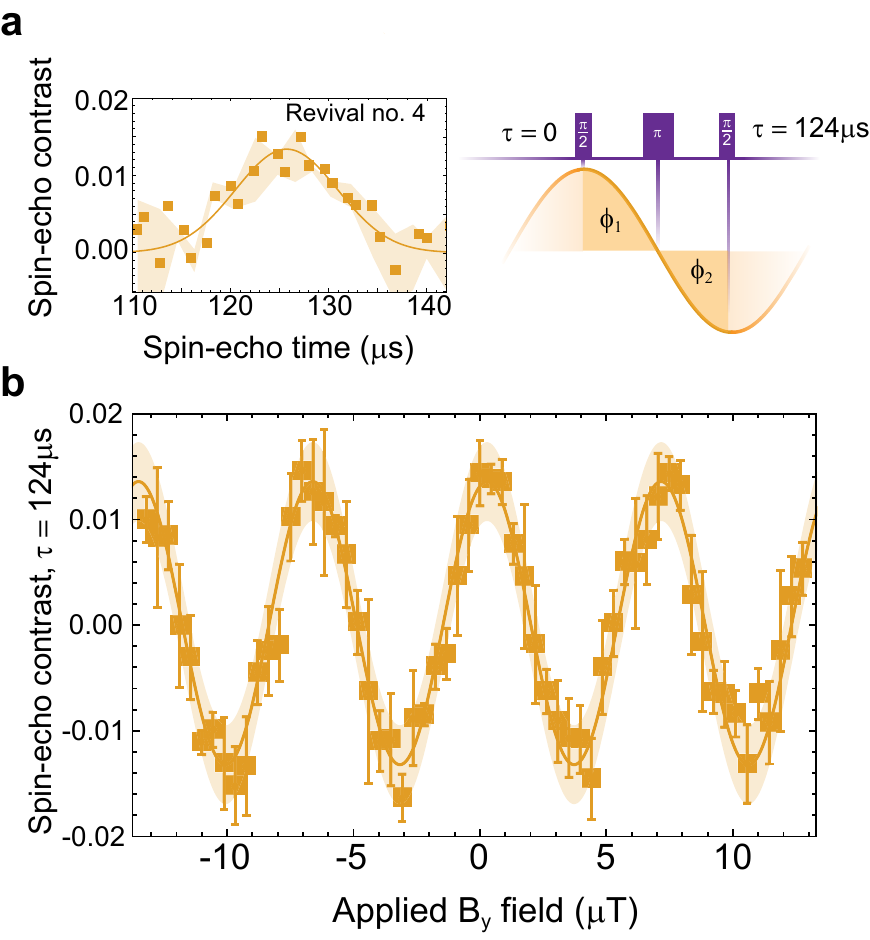}
	\caption{Rotational up-conversion magnetometry results. (a)~We use an interrogation time of $\tau = 124\,\upmu$s at a rotation speed of 3.33\,kHz, at which the fourth $^{13}$C revival provides the optimum signal to noise for AC magnetometry. The $\pi$-pulse of the echo sequence is concurrent with the zero-crossing of the Zeeman modulation from the up-converted DC field, so that equal and opposite phase shifts ($\phi_1 = -\phi_2$) are accumulated on each side of the pulse sequence, yielding the maximum sensitivity for $\tau$ less than one rotation period ($300\,\upmu$s). The statistical error bars are smaller than the data points. (b)~Spin-echo signal at $\tau = 124\,\upmu$s for an additional field applied along the $y$-axis, orthogonal to $z$. Error bars are standard deviation of three repeated measurements at each $B_y$ consisting of $2.5\times10^5$ repetitions of an echo sequence with $\pi/2$ and $3\pi/2$ readout. Lines are sinusoidal fits and shaded regions denote average error bounds.}
	\label{fig:VSWIM}
\end{figure}

 For maximum phase accumulation and hence best sensitivity, the spin-echo measurement time should be equal to the period of the up-converted field; that is, the rotation period. However, the required rotation speeds ($T_2^{-1} =7\,$kHz) cannot be sustained by our motor for the extended durations required to achieve adequate photon counting statistics. We therefore used $f_\text{rot} = 3.33\,$kHz, and adjusted the sequence timing (which is always phase synchronous with the up-converted DC field) so that each half of the spin-echo sequence measures an equal and opposite phase either side of the up-converted field zero crossing (Fig. \ref{fig:VSWIM}(a))\footnote{The maximum phase accumulation possible in this configuration is thus $2.7$ times less than if the whole period were measured}. We then varied the applied $B_y$ field by changing the current in the $y$-oriented coil pair. Figure \ref{fig:VSWIM}(b) shows the spin echo signal for a measurement time of $\tau = 124\,\upmu$s as a function of the applied $B_y$ field, confirming DC fields can be detected using rotational up-conversion. In $T_\text{int} = 300$s of measurement time, we perform $2.5\times10^5$ repetitions of both $\pi/2$ and $3\pi/2$ readouts. The minimum detectable field is given by
\begin{equation}
\delta B_\text{min} = \sigma (d\mathcal{S}/dB)^{-1},
\label{eq:dsdb}
\end{equation}
with $\sigma$ the standard deviation of the spin-echo signal $\mathcal{S}$ taken from three repeated measurements. For this data we find $\delta B_\text{min} = 0.33(2)\,\upmu$T, with a corresponding sensitivity per unit bandwidth of $\eta = B_\text{min}\sqrt{T_\text{int}} = 5.8(4)\,\upmu\text{T\,Hz}^{-1/2}$ at DC. 

\section{Comparison with standard Ramsey DC magnetometry} 
\begin{figure}
	\centering
		\includegraphics{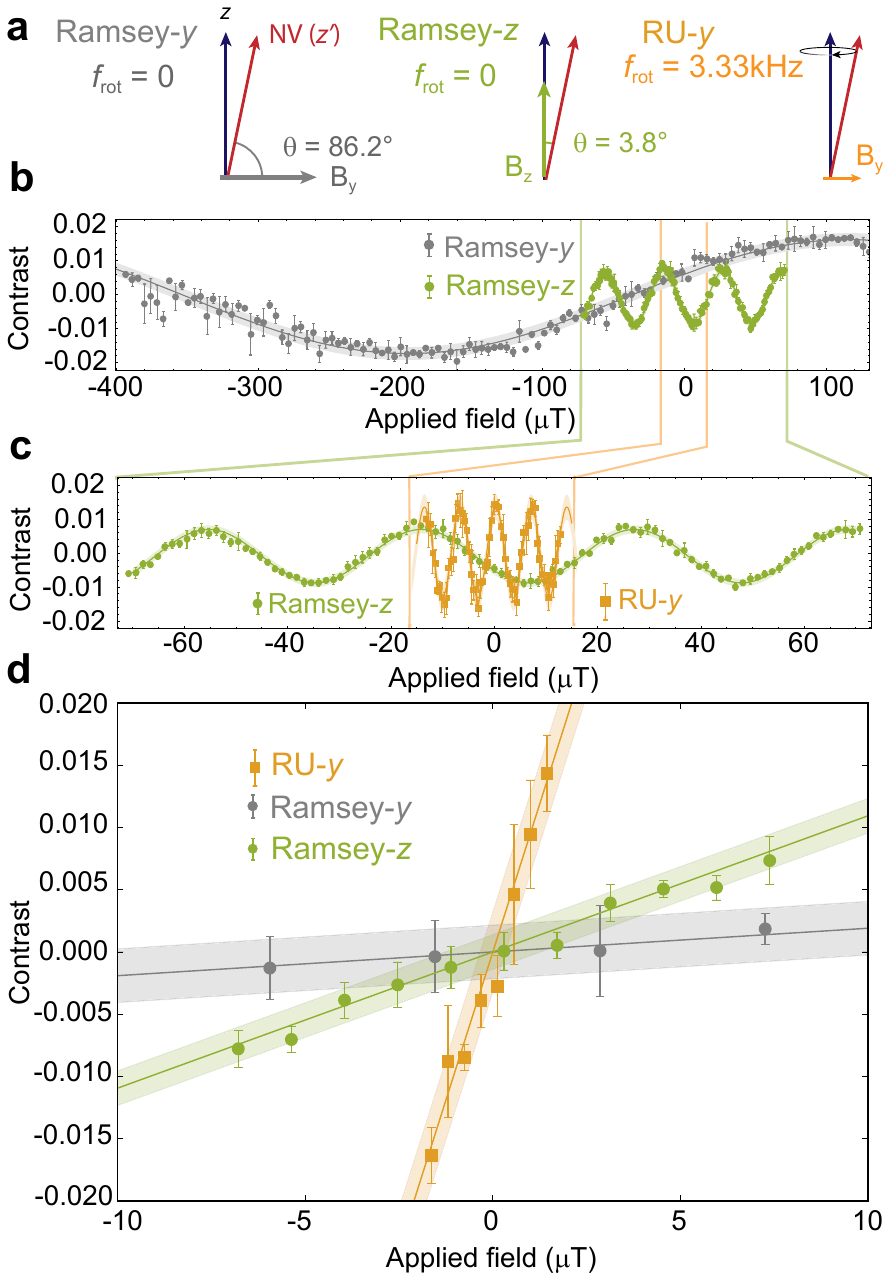}
	\caption{Comparison between rotational up-conversion and standard DC Ramsey magnetometry. (a) Configurations of NV axis and magnetic field vectors for Ramsey-$y$, Ramsey-$z$ and RU-$y$. All experiments were conducted with a $B_0 = 5.7\,$mT magnetic bias field parallel to the rotation axis. Spin-echo magnetometry signals for (b) $y$-Ramsey (grey circles) and $z$-Ramsey (green circles) compared to rotational up-conversion (orange squares) (c). All measurements consist of average signal from three separate acquisitions of $10^6$ experimental repetitions (Ramsey) and $2.5\times10^5$ repetitions (RU), error bars denote standard deviation. Lines are sinusoidal fits and shaded regions denote average error bounds. (d) Comparison of linear regions of magnetometer signal for all three techniques, with the $B$-axis scaled so that all traces intercept at $B = 0$.}
	\label{fig:senscomp}
\end{figure}

The operational sensitivity determined for DC up-conversion magnetometry is modest in comparison to other NV-based DC sensors based on ODMR or Ramsey interferometry~\cite{rondin_magnetometry_2014}. The operating sensitivity, which derives from the minimum detectable field (Eq. \ref{eq:dsdb}), depends on many factors specific to individual experiments, such as collection efficiency, state manipulation fidelity and the magnitude of noise in the environment where the sensor is evaluated. The focus of this work is the response of the NV interferometric signal to an applied magnetic field, as the process of up-conversion demonstrated here primarily increases the magnetometer response $d\mathcal{S}/dB$. For this reason, we compare the response of our rotational up-conversion magnetometry to conventional, stationary Ramsey interferometry in the same experimental environment (Figure \ref{fig:senscomp}(a)).

To compare up-conversion magnetometry to Ramsey, the diamond is first held static and oriented so that the NV axis is tilted toward the $y$-axis, ensuring maximum possible sensitivity along this axis. In this configuration (denoted Ramsey-$y$), the NV axis makes an angle of $\theta_y = 86.2^\circ$ to the $y$-field, and the Zeeman shift of the $m_S = -1$ transition is still linearly approximated by $\omega_0+B_y\,\cos(\theta_y)$ for $B_y<0.5\,$mT \footnote{See Supplemental Material at [URL] for details of model, magnetic field configuration and coordinate frame.}. While Ramsey-$y$ serves as a like-for-like comparison between upconverted and stationary measurement of a transverse field, it is not representative of the optimum DC magnetometry in our setup, which would align the test field along the NV axis. We therefore also compared the Ramsey response to a $z$-oriented field (Ramsey-$z$), which makes a much smaller angle of $3.8^\circ$ to the NV axis. Due to the smaller angle to the NV axis, the magnetometer is 15 times more sensitive to a change in the $z$-field compared to a change in the $y$-field of equal magnitude, and is essentially the best DC magnetometry possible with our experiment, $\gamma_e B_z \cos\,4^\circ\approx\gamma_e B_z$. Figure \ref{fig:senscomp}(b, c) shows the interferometry signals from rotational up-conversion sensing of a $y$-field (RU-$y$) and stationary Ramsey-$y$ and Ramsey-$z$ interferometry with $\tau  = 0.86\,\upmu$s\footnote{We use $\tau  = 0.86\,\upmu$s rather than $\tau  = T_2^*$ due to the hyperfine modulation of the Ramsey contrast, evident in Fig. \ref{fig:revivalsdata}}.

The central result is shown in Figure \ref{fig:senscomp}(d), where we compare the response of the interferometric signal to an applied magnetic field for all three techniques, and the consequent sensitivity parameters are summarised in Table \ref{tab:comp}. Rotational up-conversion offers a substantial increase in the magnetometer response $d\mathcal{S}/dB$: 50 times larger than Ramsey-$y$ and 9 times larger than Ramsey-$z$. These measurements confirm the basic premise of our technique: a longer measurement time, and thus increased sensitivity, is possible only because of elimination of the nuclear spin bath dephasing. However, the signal noise in rotational up-conversion measurements is almost twice that of Ramsey-$y$, which is in turn almost 1.6 times greater than Ramsey-$z$~\footnote{The additional noise in the Ramsey-$y$ data is attributed to lower stability of the coil current power supply at higher currents.}. Combined with the longer integration time of the up-conversion measurements compared to the Ramsey measurements, this leads to a comparable sensitivity per-unit-bandwidth for up-conversion and Ramsey-$z$ in the current setup with low $\theta_\text{NV}$. 

Some increased noise of the up-converted DC signal is expected, because roughly four times fewer photons are collected in the 300\,s integration time used for rotational upconversion compared to the shorter Ramsey experiments, which run more repetitions in the 10\,s integration time. We observe that the Ramsey measurements exhibit photon collection statistics only $0.6$\,dB above the shot noise limit, while with rotational up-conversion we measure photons at 3.6\,dB above photon shot noise, which we attribute to variations in the laser power over the longer integration times of RU-$y$. The increased dead-time of the rotational up-conversion measurement also plays a significant role. In our experiments, the spin-echo measurement time $\tau$ is dictated by the coherence time of the diamond (and $^{13}$C revival time), but the duty cycle is determined by the rotation speed of the motor. The finite pumping time of the laser readout pulse impinges into the next period of the rotation, making back-to-back $\pi/2$ and $3\pi/2$ readouts on alternate periods of the motor difficult with our current experimental hardware. This effectively limits the duty cycle $20\,\%$, one $124\,\upmu$s spin echo measurement every \emph{two} $300\,\upmu$s rotation periods. This effect alone results in comparable per-unit-bandwidth sensitivities for static and rotating measurements. 

\section{Noise sources in up-conversion magnetometry}

It is interesting to examine the sensitivity results from the previous Section in more detail, since it suggests that either the up-conversion method is imperfect, or that a significant amount of noise is being up-converted. We will now briefly discuss the importance of several noise sources particular to rotational upconversion, such as specific technical issues, upconversion of off-axis noise and imperfect rotation. 

In general, we can subdivide the noise sources that perturb the two-level splitting into in-diamond \emph{instrinsic} sensor noise sources that co-rotate with the NV sensor, such as the surrounding $^{13}$C nuclear spins, and \emph{extrinsic} environmental noise in the stationary lab frame, such as drifting magnetic bias fields or temperature variations. Since the diamond rotates around a particular axis in space, the up-conversion or suppression of magnetic field noise is vectorial in nature. In contrast, temperature shifts are rotationally invariant and not modulated by sensor rotation, and are therefore eliminated by the spin-echo sequence. Environmental magnetic field drifts are either upconverted to the rotation frequency if transverse to the rotation axis, or eliminated by the spin-echo sequence if parallel to it. The latter effect is an added benefit to using a $z-$oriented magnetic bias field: drifts in the amplitude of the bias field parallel to the rotation axis will not be upconverted to AC, since such drifts appear as static level shifts in a spin-echo sequence. Noise or drifts in DC field components transverse to the rotation axis will be upconverted to AC. 

A spin-echo sequence applied to a stationary NV refocuses the coherent in-diamond bath noise as well as all quasi-static DC field perturbations. In the rotating up-conversion sequence, the same components of bath noise are refocused, but only one vector component (parallel to $z$, the rotation axis) of the extrinsic magnetic noise is suppressed. An associated benefit is that intrinsic bath noise, which is typically the dominant contributor to the ensemble dephasing time $T_2^\ast$, can then be separated from noise, or signal, in the sensing environment. On the other hand, up-converted noise may result in a reduced signal-to-noise ratio of the DC field of interest. The process of up-conversion, in our case linked to the performance of the electric motor, may also introduce noise. For example, wobbling of the motor spindle or a jittering rotation period will introduce noise into the upconverted signal. We have recently used identical apparatus to rotate diamonds containing single NV centres~\cite{wood_quantum_2018} that could still be reliably imaged near the diffraction limit and controlled at up to $f_\text{rot} = 5.2\,$kHz, suggesting the mechanical rotation is of high enough quality to be ruled out as a significant contribution to noise in the up-converted signal. 

 Other experiments have detected evidence of drifts and current ripple in the bias coils used to create magnetic fields, which would have a proportionately more severe effect on rotational up-conversion. Fast current noise originating from power supply switching (up to $\sim100\,$kHz)  will be detectable in a spin-echo measurmement, rotating or stationary, but not Ramsey with $\tau<1\,\upmu$s. Further work, especially testing in a significantly cleaner magnetic environment, is needed to conclusively identify the role of noise sources in rotational up-conversion when compared to the Ramsey measurement in this work.  

\section{Operating sensitivity\label{sec:sens}}

The shot-noise limited sensitivity for rotational up-conversion magnetometry is approximated by 
\begin{equation}
\eta \approx \frac{\pi}{\gamma 2C\,\sin\theta_\text{NV}}\frac{\sqrt{\tau+t_D}}{\tau}
\label{eq:sss}
\end{equation}
with $C= 0.02$ the state readout efficiency, which depends on the photon detection efficiency, the NV density, the signal contrast and the photon emission rate per NV ~\cite{rondin_magnetometry_2014,taylor_high-sensitivity_2008, degen_quantum_2017}. The dead time $t_D$ includes the time required to prepare and readout the NV spin. Table \ref{tab:comp} summarizes the operational and projected shot-noise limited sensitivity for each technique compared in this work with our current experimental parameters and limitations. We note that the duty cycle limitations in our current realisation (detailed in the previous Section) result in a comparable shot-noise limited sensitivity for Ramsey and up-conversion magnetometry, a factor of 2 below what we observed. With these limitations in mind, we can also calculate the potential of rotational up-conversion with our demonstrated rotation speeds and collection efficiencies but using a diamond sample with two simple alterations: a larger angle $\theta_\text{NV} = 54.7^\circ$, which is obtained by using a diamond with a $(100)$-cut face and a slightly longer $T_2$ time so that we may measure for a full period of the up-converted DC field, $\tau = 300\,\upmu$s. These simple, experimentally feasible~\cite{stanwix_coherence_2010} improvements highlight the potential of rotational up-conversion, and would lead to a 32 times improvement over the proof-of-principle results demonstrated in this work. Additional improvements to DC field sensitivity are detailed in the Discussion.

\begin{table}
\begin{ruledtabular}
\begin{tabular}{lcccc}
   & $d\mathcal{S}/dB$ & $\eta_\text{Opr}$ & $\eta_\text{SN}$ \vspace{1.2mm}\\

units  &  $10^{-3}\,\upmu$T$^{-1}$& $\upmu\text{T\,Hz}^{-1/2}$ & $\upmu\text{T\,Hz}^{-1/2}$  \\
\hline
Ramsey-$y$  &  0.02(1)& 35(9) & 25 \\
Ramsey-$z$  &  1.1(1)& 4.0(2) & 1.8 \\
RU-$y$   &  9.5(6)& 5.8(4) & 2.3  \\
RU-$y$ (best)   &  308 & - & 0.08  \\

\end{tabular}
\end{ruledtabular}
\caption{DC sensitivity comparisons for the three experimentally demonstrated techniques and rotational up-conversion with two simple alterations: measuring for the whole period of the up-converted field ($\tau = 300\,\upmu$s) and using a $(100)$-cut diamond with $\theta_\text{NV} = 54.7^\circ$. The parameters are response $d\mathcal{S}/dB$, operating sensitivity per unit bandwidth $\eta_\text{Opr}$, shot-noise limited sensitivity with operating duty cycle $\eta_\text{SN}$.}
\label{tab:comp}
\end{table}

\section{Discussion}  
The purpose of this work is to introduce and demonstrate the feasibility of a new method of magnetic sensing with NV spin sensors. Rotational up-conversion magnetometry in a proof-of-principle stage exhibits the increased sensing time and concomitant improved response to DC fields compared to conventional Ramsey sensing, and in a like-for-like comparison (RU-$y$ and Ramsey-$y$) exhibits a substantially improved sensitivity per unit bandwidth. It is worth noting that although not representative of the limits of either up-conversion sensing or DC magnetometry using Ramsey interferometry, the improvement to \emph{transverse} field sensing demonstrated herein is indicative of the benefits to DC sensing our technique promises. Technical limitations, principally the low $\theta_\text{NV}$, preclude a sensitivity-per-unit bandwidth comparable to optimised AC magnetic sensing experiments, or exceeding the best Ramsey magnetometry possible in our setup (Ramsey-$z$). Ref.~\cite{taylor_high-sensitivity_2008} provides a comprehensive prescription for improving the magnetometer sensitivity, and Eq. (\ref{eq:sss}) suggests several simple avenues that could improve the sensitivity of rotational up-conversion to that demonstrated with AC field sensing experiments. The experimental duty cycle can be made near unity for coherence times equal to or exceeding the rotation period of the diamond, and the need to operate on every second rotation period can be circumvented with a retriggerable pulse generator.   

Diamonds with a higher NV density could be used, though a tradeoff exists between the density of the NV ensemble and the resulting coherence time: high density, nitrogen-rich samples ($N\gtrsim 1$\,ppm) increase the number of participating NV sensors, and thus photon collection up to a point beyond which $T_2$ is reduced due to interactions with paramagnetic nitrogen centres~\cite{taylor_high-sensitivity_2008}. For rotational upconversion to be effective, the minimum requirement is that $1/f_\text{max}\lesssim T_2$. For this reason, we consider ensemble densities where NV-P1 interactions are negligible ($n_\text{NV}\lesssim 10^{15}\,\text{cm}^{-3})$ and coherence times are on the order of rotational periods achievable with our current motor. Previous work has demonstrated ensembles with natural abundance $^{13}$C to have coherence times of $T_2\approx600\,\upmu$s~\cite{stanwix_coherence_2010}, and isotopically-pure $^{12}$C diamonds hosting single NV centres have been found to have coherence times of greater than $2\,$ms~\cite{balasubramanian_ultralong_2009}, still substantially higher than the corresponding $T_2^\ast$ dephasing times of $\sim100\,\upmu$s. Such long coherence times offer the prospect of measuring multiple rotations with the speeds possible in our experiments. For a diamond with $T_2 = 2\,$ms and the maximum rotation speed of the motor ($8.3\,$kHz), almost 17 complete rotations could be observed, and multiple-pulse sequences such as XY-N or CPMG~\cite{degen_quantum_2017} could be employed, resulting in a signficant improvement in sensitivity.

The angle $\theta_\text{NV}$ between the NV axis and the rotation axis can also be increased; for example, a $(100)$-cut diamond offers $\theta_\text{NV} = 54.7^\circ$.  As discussed in in the Experiment section, our demonstration used a $(111)$-cut diamond with small $\theta_\text{NV}$ to easily allow a single orientation class to be isolated and addressed, this issue can be resolved by using diamonds containing NVs with a high degree of preferential orientation ($>90\%$)~\cite{michl_perfect_2014}. Diamonds containing all four orientation classes could also be laser-cut and polished to a preferred angle, allowing for four independently addressable NV orientation classes, which would enable vector sensitivity~\cite{steinert_high_2010} in the rotational plane. Alternatively, 2D vector sensitivity could be acheived by measuring a single orientation class at rotation angles separated by $90^\circ$ on alternate rotation cycles, effectively sampling the up-converted field modulation in quadrature. Using preferentially-oriented samples would increase the contrast of the spin-echo signal, since less non-participating NV centres would be present~\cite{pham_enhanced_2012}. A diamond with an ensemble density of NV centres peferentially aligned along one axis with the same photon count rate as the diamond used in this work ($3\times10^6\,\text{s}^{-1}$) would yield an order of magnitude improvement in state readout efficiency, $C = 0.1$. Additionally, using an $n = 17$ $\pi$-pulse sequence at a rotation speed of $8.3\,$kHz, $C =  0.1$, $\theta_\text{NV} = 54.7^\circ$ and $T_2 = 2\,$ms results in a DC shot-noise limited sensitivity of 0.3\,nT\,Hz$^{-1/2}$. This projected sensitivity relates to the same experimental configuration described in this work, only with a different diamond sample. 

More substantial improvements to photon collection efficiency by using a larger optical addressing region~\cite{wolf_subpicotesla_2015,clevenson_broadband_2015} could allow DC field sensing into the picotesla range, allowing realistic diamond-based quantum sensors to access improved sensitivity to static magnetic fields, and with the immunity to drifts in the ambient temperature provided by spin-echo interferometry. Ultimately, up-conversion offers the prospect of a maximum $\sqrt{T_2^\ast/T_2}$ improvement in sensitivity. Our demonstration here focuses on magnetic fields, but equivalent improvements to static electric field sensing could be possible, where the direction of the electric field (and strain field) determines the level splitting~\cite{dolde_electric-field_2011, doherty_measuring_2014}. Another interesting extension concerns geometric phase accumulated by the NV as its axis is rotated~\cite{maclaurin_measurable_2012}. Although negligible in this work, in the proposed improvements where the NV axis makes a significant angle to the rotation axis, geometric phase accumulation is substantial, up to $2.7\,$rad for a complete rotation with $\theta_\text{NV} = 54.7^\circ$. We note however that the geometric phase is otherwise an easily accounted for level shift that can be cancelled in a spin-echo sequence. Alternatively, measurement of geometric phase can serve as a gyroscope~\cite{ledbetter_gyroscopes_2012}, an independent diagnostic of the stability of rotation as a means of DC field upconversion.  

The barriers to achieving superlative magnetic sensitivity with our method are common to many NV magnetometry schemes, as the projected sensitivity of our magnetometer is impacted by fluctuating or drifting magnetic bias fields, local magnetic field gradients and local temperature gradients and drifts. Temperature-induced shifts are not modulated by rotation, and so may be eliminated by the spin echo sequence. The impact of local magnetic field gradients, a problem which scales with the size of the optical addressing region, is an issue in most magnetometers, i.e. as the size of the sensing volume increases so does the impact of field gradients, though with optimised design of magnetic bias fields can be rendered manageably small \cite{bauch_ultralong_2018}. 
  
\begin{acknowledgments}
We acknowledge valuable discussions with L. P. McGuinness, L. T. Hall, D. A. Simpson and J.-P. Tetienne. A.~M.~M. thanks the Institute of Advanced Study (Durham University, U.K.) for hosting him during the preparation of this manuscript. This work was supported by the Australian Research Council Discovery Scheme (DP150101704). 
\end{acknowledgments}

\end{document}